\documentclass{amsproc}

\usepackage{multicol,amssymb,amsmath,amsbsy,amsthm,graphicx,esvect}
\usepackage[latin1]{inputenc}
\usepackage{caption2}

\newtheorem{theorem}{Theorem}[section]
\newtheorem{lemma}[theorem]{Lemma}
\newtheorem{proposition}[theorem]{Proposition}
\newtheorem{corollary}[theorem]{Corollary}

\theoremstyle{definition}
\newtheorem{definition}[theorem]{Definition}

\theoremstyle{remark}
\newtheorem{remark}[theorem]{Remark}

\numberwithin{equation}{section}

\newcommand{\supp}{\text{\rm{supp}}}
\newcommand{\card}{\text{\rm{card}}}

\newcommand{\capp}{\text{\rm{cap}}}

\begin{document}

\title{Greedy energy points with external fields}

\author{A. L\'{o}pez Garc\'{i}a}
\address{Department of Mathematics, Vanderbilt University, Nashville TN 37240, USA}

\email{abey.lopez@vanderbilt.edu}


\dedicatory{Dedicated to my father, G. L\'{o}pez Lagomasino, on
the occasion of his 60th birthday.}

\keywords{Minimal energy, Leja points, Equilibrium measure,
External field, Riesz kernels, One-point compactification.
\\ \indent The results of this paper form a part of this author's Ph.D. dissertation
at Vanderbilt University, under the direction of Edward B. Saff.
Research was supported in part by the U.S. National Science
Foundation grant DMS-0808093.}

\begin{abstract}
In this paper we introduce several extremal sequences of points on
locally compact metric spaces and study their asymptotic
properties. These sequences are defined through a greedy algorithm
by minimizing a certain energy functional whose expression
involves an external field. Some results are also obtained in the
context of Euclidian spaces $\mathbb{R}^{p}$, $p\geq 2$. As a
particular example, given a closed set $A\subset\mathbb{R}^{p}$, a
lower semicontinuous function
$f:\mathbb{R}^{p}\rightarrow(-\infty,+\infty]$ and an integer
$m\geq 2$, we investigate (under suitable conditions on $A$ and
$f$) sequences $\{a_{i}\}_{1}^{\infty}\subset A$ that are
constructed inductively by selecting the first $m$ points
$a_{1},\ldots,a_{m}$ so that the functional
\[
\sum_{1\leq i<j\leq
m}\frac{1}{|x_{i}-x_{j}|^{s}}+(m-1)\sum_{i=1}^{m}f(x_{i})
\]
attains its minimum on $A^{m}$ for $x_{i}=a_{i}$, $1\leq i\leq m$,
and for every integer $N\geq 1$, the points
$a_{mN+1},\ldots,a_{m(N+1)}$ are chosen to minimize the expression
\[
\sum_{i=1}^{m}\sum_{l=1}^{mN}\frac{1}{|x_{i}-a_{l}|^{s}}
+\sum_{1\leq i<j\leq
m}\frac{1}{|x_{i}-x_{j}|^{s}}+((N+1)m-1)\sum_{i=1}^{m}f(x_{i})
\]
on $A^{m}$. We assume here that $s\in[p-2,p)$. An extension of a
result due to G. Choquet concerning point configurations with
minimal energy is obtained in the context of locally compact
metric spaces and constitutes a key ingredient in our analysis.
\end{abstract}

\maketitle

\section{Introduction}\label{Introduction}

In this paper we study asymptotic properties of certain extremal
sequences of points defined on locally compact metric spaces. We
shall refer to them as \textit{greedy energy sequences}. This
terminology was recently introduced in \cite{LS}. These sequences
are indeed generated by means of a greedy algorithm at every step
of which a certain energy expression is minimized. The notion of
energy that we refer to will be specified shortly. The asymptotic
properties that we analyze are mainly the following: if
$\{\alpha_{N}\}_{N}$ denotes the sequence of configurations formed
by the first $N$ points of a greedy energy sequence, we use
potential-theoretic tools to study the behavior of the energy of
$\alpha_{N}$ as $N$ approaches infinity and the limiting
distributions of these configurations. We remark that in \cite{LS}
a number of results about greedy sequences were obtained in a
context in which potential theory is no longer applicable.

Potential theory on locally compact Hausdorff (LCH) spaces is a
classical field which was developed, among others, by Choquet
\cite{Choquet2, Choquet}, Fuglede \cite{Fuglede} and Ohtsuka
\cite{Ohtsuka}. In recent years, and also in the context of LCH
spaces, Zorii \cite{Zorii, Zorii2} has studied solvability
properties of the Gauss variational problem in the presence of an
external field. A similar problem (we shall also call it Gauss
variational problem) is considered below. We next introduce the
basic notions necessary to describe our results.

Let $X$ denote a locally compact metric space containing
infinitely many points. If $X$ is not compact, let
$X^{*}=X\cup\{\infty\}$ denote the one-point compactification of
$X$. A \textit{kernel} in $X$ is, by definition, a lower
semicontinuous function (l.s.c.) $k:X \times X \rightarrow
\mathbb{R}\cup\{+\infty\}$. It is called \textit{positive} if
$k(x,y)\geq 0$ for all $x, y\in X$.

Assume that $f:X\longrightarrow\mathbb{R}\cup\{+\infty\}$ is a
l.s.c. function. For a set $\omega_N=\{x_1,\ldots,x_N\}$ of $N$
($N\geq 2$) points in $X$ which are not necessarily distinct, we
write $\card(\omega_{N})=N$ and define the \textit{energy} of
$\omega_N$ by
\begin{equation*}
E(\omega_N):=\sum_{1\leq i\neq j\leq N}k(x_{i},x_{j})
=\sum_{i=1}^{N}\sum_{j=1,j\neq i}^{N}k(x_{i},x_{j})\,,
\end{equation*}
whereas the \textit{weighted energy} of $\omega_{N}$ is given by
\begin{equation*}
E_{f}(\omega_{N}):=E(\omega_N) +2(N-1)\sum_{i=1}^{N}f(x_{i})\,.
\end{equation*}
In potential theory the function $f$ is usually referred to as an
\textit{external field}.

If the kernel is \textit{symmetric}, i.e., $k(x,y)=k(y,x)$ for all
$x, y\in X$, we may also write
\begin{equation*}
E(\omega_N)=2 \sum_{1\leq i < j\leq N}k(x_{i},x_{j})\,.
\end{equation*}

\begin{definition}
For a non-empty set $A\subset X$, the \textit{weighted $N$-point
energy} of $A$ is given by
\begin{equation}\label{defnminimalenergweighted}
\mathcal{E}_{f}(A,N):=\inf\{E_{f}(\omega_{N}):\omega_N \subset A,
\,\,\card(\omega_N)=N\}\,.
\end{equation}
In case that $f\equiv 0$, we use instead the notation
\begin{equation}\label{defnminimalenerg}
\mathcal{E}(A,N):=\inf\{E(\omega_{N}):\omega_N \subset A,
\,\,\card(\omega_N)=N\}\,.
\end{equation}
We say that $\omega_{N}^{*}\subset A$ is an \textit{optimal
weighted $N$-point configuration} on $A$ if
\begin{equation*}
E_{f}(\omega_{N}^{*})=\mathcal{E}_{f}(A,N)\,.
\end{equation*}
If $A$ is compact, the existence of $\omega_{N}^{*}$ follows from
the lower semicontinuity of $k$ and $f$.
\end{definition}

It is necessary to introduce now the continuous counterparts of
the above notions. Let $\mathcal{M}(A)$ denote the linear space of
all real-valued Radon measures that are compactly supported on
$A\subset X$, and let
\begin{equation*}
\mathcal{M}^{+}(A):=\{\mu\in\mathcal{M}(A): \mu\geq
0\}\,,\qquad\mathcal{M}_{1}(A):=\{\mu\in\mathcal{M}^{+}(A):
\mu(X)=1\}\,.
\end{equation*}
Given a measure $\mu\in\mathcal{M}(X)$, the \textit{energy} of
$\mu$ is the double integral
\begin{equation}\label{definicionenergia}
W(\mu):=\int\int k(x,y)\,d\mu(x)\,d\mu(y)\,,
\end{equation}
whereas the function
\begin{equation}\label{defnpotential}
U^{\mu}(x):=\int k(x,y) \,d\mu(y)
\end{equation}
is called the \textit{potential} of $\mu$. The \textit{weighted
energy} of $\mu$ is defined by
\begin{equation}\label{defweightedenergy}
I_{f}(\mu):=W(\mu)+2\int f \,d\mu\,.
\end{equation}
Since any l.s.c. function is bounded below on compact sets, the
above integrals are well-defined, although they may attain the
value $+\infty$.

The quantity $w(A):=\inf\{W(\mu):\mu\in\mathcal{M}_{1}(A)\}$ is
called the \textit{Wiener energy} of $A$, and plays an important
role in potential theory. The \textit{capacity} of $A$ is defined
as $\rm{cap}$$(A):=w(A)^{-1}$ if $k$ is positive, and otherwise,
it is defined as $\rm{cap}$$(A):=\exp(-w(A))$. A property is said
to hold \textit{quasi-everywhere} (q.e.), if the exceptional set
has Wiener energy $+\infty$.

Given a net $\{\mu_{\alpha}\}\subset\mathcal{M}(A)$, we say that
$\{\mu_{\alpha}\}$ converges in the \textit{weak-star topology} to
a measure $\mu\in\mathcal{M}(A)$ when
\begin{equation*}\lim_{\alpha}\int g \,d\mu_{\alpha}=\int g \,d\mu\,,
\qquad \mbox{for all}\quad g\in C_{c}(A)\,,\end{equation*} where
$C_{c}(A)$ denotes the space of compactly supported continuous
functions on $A$. We will use the notation
\begin{equation*}
\mu_{\alpha}\stackrel{*}{\longrightarrow}\mu \end{equation*} to
denote the weak-star convergence of measures.

If $w(A)<\infty$, a measure $\mu\in\mathcal{M}_{1}(A)$ satisfying
the property $W(\mu)=w(A)$ is called an \textit{equilibrium
measure}. If $A$ is compact, the existence of such a measure is
guaranteed by the lower semicontinuity of $k$ and the compactness
of $\mathcal{M}_{1}(A)$ equipped with the weak-star topology (cf.
\cite[Theorem 2.3]{Fuglede}). However, uniqueness does not always
hold.

The following result is due to G. Choquet \cite{Choquet}, and it
is central in this theory.
\begin{theorem}\label{Choquettheo}
Let $k$ be an arbitrary kernel and $A\subset X$ be a compact set.
If $\{\omega_{N}^{*}\}$ is a sequence of optimal $N$-point
configurations on $A$, then
\begin{equation*}
\lim_{N\rightarrow\infty}\frac{E(\omega_{N}^{*})}{N^{2}}=w(A)\,.
\end{equation*}
\end{theorem}
The following variation of Theorem \ref{Choquettheo} was obtained
by B. Farkas and B. Nagy \cite{FarkasNagy}.
\begin{theorem}\label{FNtheo}
Assume that the kernel $k$ is positive and is finite on the
diagonal, i.e., $k(x,x)<+\infty$ for all $x\in X$. Then for
arbitrary sets $A\subset X$,
\begin{equation*}
\lim_{N\rightarrow\infty}\frac{\mathcal{E}(A,N)}{N^{2}}=w(A)\,,
\end{equation*}
where $\mathcal{E}(A,N)$ is defined by $(\ref{defnminimalenerg})$.
\end{theorem}

We remark that Theorems \ref{Choquettheo} and \ref{FNtheo} were
proved in the context of LCH spaces, but for the sake of
uniformity we always assume, unless otherwise stated, that $X$
denotes a locally compact metric space.

In this paper we are interested in the so-called \textit{Gauss
variational problem} in the presence of an external field $f$. In
what follows we assume that $A\subset X$ is a closed set, and we
will refer to $A$ as the \textit{conductor}. The Gauss v.p.
consists of finding a solution to the minimization problem
\begin{equation}\label{Gaussvarprob}
V_{f}(A):=\inf_{\mu\in\mathcal{M}_{f}(A)}I_{f}(\mu)\,,
\end{equation}
where $\mathcal{M}_{f}(A)$ denotes the class of measures
\begin{equation}\label{classmeasuresextfield}
\mathcal{M}_{f}(A):=\{\mu\in\mathcal{M}_{1}(A): W(\mu), \int f
\,d\mu < +\infty\}\,.
\end{equation}
Throughout the rest of the paper we will denote $V_{f}(A)$ simply
as $V_{f}$. If $\mathcal{M}_{f}(A)=\emptyset$ then by definition
$V_{f}=+\infty$. If $\mathcal{M}_{f}(A)\neq\emptyset$ and there
exists a minimizing measure $\mu\in\mathcal{M}_{f}(A)$ satisfying
$I_{f}(\mu)=V_{f}$, we call $\mu$ an \textit{equilibrium measure
in the presence of the external field} $f$. In this case we say
that the Gauss variational problem is \textit{solvable}, and
observe that $V_{f}$ is finite.

Sufficient conditions for the existence and uniqueness of solution
for a similar variational problem were provided by N. Zorii (see
\cite{Zorii,Zorii2}) in the more general context of LCH spaces.
She assumes that the kernel is positive if $A$ is not compact, and
allows measures to have non-compact support in this case. We
remark that the theory of logarithmic potentials ($k(x,y)=-\log
|x-y|$) with external fields in the complex plane is particularly
rich in applications to physics and other branches of analysis. We
refer the reader to \cite{SaffTotik} for details on this theory.

Let us introduce the notation
\begin{equation}\label{eq:defnWf}
W_{f}(\mu):=V_{f}-\int f \,d\mu
\end{equation}
for an equilibrium measure $\mu\in\mathcal{M}_{f}(A)$. This value
is finite. The \textit{essential support} of $\mu$ is defined as
\begin{equation*}
S_{\mu}^{*}:=\{x\in A: U^{\mu}(x)+f(x)\leq W_{f}(\mu)\}\,.
\end{equation*}
Using Lemma 2.3.3 from \cite{Fuglede} and the argument employed in
\cite{SaffTotik} to prove parts $(d)$ and $(e)$ of Theorem I.1.3,
it is easy to see that if $k$ is a symmetric kernel and
$\mu\in\mathcal{M}_{f}(A)$ is an equilibrium measure, then
\begin{equation}\label{Gaussvarineq1}
U^{\mu}(x)+f(x)\leq V_{f}-\int f \,d\mu
\end{equation}
holds for all $x\in\supp(\mu)$ (i.e. $\supp(\mu)\subset
S_{\mu}^{*}$) and
\begin{equation}\label{Gaussvarineq2}
U^{\mu}(x)+f(x)\geq V_{f}-\int f \,d\mu
\end{equation}
holds q.e. on $A$.

We are ready to introduce the following definitions (see also
Definitions $\ref{defngreedyprimera2}$ and
$\ref{defngreedypointsclosed2}$ in Section \ref{statement}).

\begin{definition}\label{defngreedyprimera}
Let $k:X \times X \rightarrow \mathbb{R}\cup\{+\infty\}$ be a
symmetric kernel on a locally compact metric space $X$, $A\subset
X$ be a closed set, and $f:X \rightarrow
\mathbb{R}\cup\{+\infty\}$ be an external field. If $X$ is not
compact, we assume that $f$ satisfies the following `growth'
condition at infinity: for each compactly supported probability
measure $\nu$,
\begin{equation}\label{condkf}
\lim_{x\rightarrow\infty}(U^{\nu}(x)+f(x))=+\infty\,,
\end{equation}
(i.e. given $M>0$, there exists a compact set $B\subset X$ such
that $U^{\nu}(x)+f(x)>M$ for all $x\in X\setminus B$).

Assume that the Gauss variational problem is solvable and
$\mu\in\mathcal{M}_{f}(A)$ is an equilibrium measure. A sequence
$(a_{n}=a_{n,f,\mu})_{n=1}^{\infty}\subset A$ is called a
\textit{weighted greedy $(f,\mu)$-energy sequence} on $A$ if it is
generated in the following way:
\begin{itemize}
\item $a_{1}$ is selected arbitrarily on $S_{\mu}^{*}$.

\item For every $n\geq 1$, assuming that $a_{1},\ldots,a_{n}$ have
been selected, $a_{n+1}$ is chosen so that $a_{n+1}\in
S_{\mu}^{*}$ and
\begin{equation}\label{defngreedypoints}
\sum_{i=1}^{n}k(a_{n+1},a_{i})+nf(a_{n+1})= \inf_{x\in
S_{\mu}^{*}}\Big\{\sum_{i=1}^{n}k(x,a_{i})+nf(x)\Big\}\,.
\end{equation}
\end{itemize}
The set formed by the first $N$ points of this sequence is denoted
by $\alpha_{N,\mu}^{f}$. We also introduce the following
associated function:
\begin{equation*}
U_{n}^{f}(x):=\sum_{i=1}^{n-1}k(x,a_{i}) +(n-1)f(x)\,,\qquad x\in
A\,,\quad n\geq 2\,.
\end{equation*}
\end{definition}

\begin{remark}
Condition $(\ref{condkf})$ implies in particular that
$S_{\mu}^{*}$ is compact. Consequently, for every $n\geq 1$, the
existence of $a_{n+1}$ is guaranteed by the lower semicontinuity
of $k$ and $f$. However, $a_{n+1}$ may not be unique.
\end{remark}

In many practical circumstances it is not possible to determine
the support or essential support of an equilibrium measure. For
this reason it is of interest to introduce the following:

\begin{definition}\label{defngreedypointsclosed}
Let $k:X \times X \rightarrow \mathbb{R}\cup\{+\infty\}$ be a
symmetric kernel on a locally compact metric space $X$, $A\subset
X$ be a closed set, and $f:X \rightarrow
\mathbb{R}\cup\{+\infty\}$ be an external field. In case it
exists, a sequence $(a_{n}=a_{n,f})_{n=1}^{\infty}\subset A$ is
called a \textit{weighted greedy $f$-energy sequence} on $A$ if it
is constructed inductively by selecting $a_{1}$ arbitrarily on $A$
such that $f(a_{1})<+\infty$, and $a_{n+1}$ as in
$(\ref{defngreedypoints})$ but taking the infimum on $A$. We use
the notation $\alpha_{N}^{f}$ to indicate the configuration formed
by the first $N$ points of this sequence.
\end{definition}

It seems that A. Edrei was the first to study in \cite{Edrei}
properties of the configurations $\alpha_{N}^{f}$ under the
assumptions $X=\mathbb{R}^{2}$, $A\subset\mathbb{R}^{2}$ is
compact, $k(x,y)=-\log |x-y|$ and $f\equiv 0$. However, in the
literature these configurations are often called \textit{Leja
points} in recognition of Leja's article \cite{Leja}.

A very important class of kernels is the so-called \textit{M.
Riesz kernels} in $X=\mathbb{R}^{p}$, which depend on a parameter
$s\in[0,+\infty)$. It is defined as follows:
\begin{equation}\label{defnRieszkernel}
k_{s}(x,y):=K(|x-y|;s)\,,\qquad x,y\in\mathbb{R}^{p}\,,
\end{equation}
where $|\cdot|$ denotes the Euclidean norm and
\begin{equation*}
K(t;s):=\left\{
\begin{array}{ccc}
t^{-s}, & \mbox{if} & s>0\,, \\
-\log(t), & \mbox{if} & s=0\,.
\end{array}
\right.
\end{equation*}

We shall use the notations $I_{s}(\mu)$, $I_{s,f}(\mu)$ and
$U_{s}^{\mu}$ to denote, respectively, the energy
(\ref{definicionenergia}), weighted energy
(\ref{defweightedenergy}) and potential (\ref{defnpotential}) of a
measure $\mu\in\mathcal{M}(\mathbb{R}^{p})$ with respect to the
Riesz $s$-kernel. We will also use the symbols $w_{s}(A)$ and
$\rm{cap}$$_{s}(A)$ to denote the Wiener $s$-energy and
$s$-capacity of a set $A\subset\mathbb{R}^{p}$ in this setting.

The paper is organized as follows. In Section \ref{statement} we
present our results and in Section \ref{Proofs} we provide their
proofs. In Section \ref{numerical} we present some numerical
computations.

\section{Statement of results}\label{statement}

Our first result is the following generalization of Theorem
\ref{Choquettheo}.
\begin{theorem}\label{primer}
Let $k:X \times X \rightarrow \mathbb{R}\cup\{+\infty\}$ be an
arbitrary kernel on a locally compact metric space $X$, $A\subset
X$ be a compact conductor, and $f:X \rightarrow
\mathbb{R}\cup\{+\infty\}$ be an external field. Assume that the
Gauss variational problem is solvable. If $\{\omega_{N}^{*}\}$ is
a sequence of optimal weighted $N$-point configurations on $A$,
then
\begin{equation}\label{asympweightmin}
\lim_{N\rightarrow\infty}\frac{E_{f}(\omega_{N}^{*})}{N^{2}}=
V_{f}\,.
\end{equation}
Furthermore, if the Gauss variational problem has a unique
solution $\mu\in\mathcal{M}_{f}(A)$, then
\begin{equation}\label{distribweightmin}
\frac{1}{N}\sum_{x\in\omega_{N}^{*}}\delta_{x}\stackrel{*}{\longrightarrow}
\mu\,,\qquad N\rightarrow\infty\,,
\end{equation}
where $\delta_{x}$ is the unit Dirac measure concentrated at $x$.
\end{theorem}

\begin{remark} As the proof of Theorem \ref{primer} shows, without assuming
the uniqueness of the equilibrium measure one can deduce that any
convergent subsequence of
$(1/N)\sum_{x\in\omega_{N}^{*}}\delta_{x}$ converges weak-star to
an equilibrium measure. This observation is also applicable to the
following result concerning greedy configurations.
\end{remark}

\begin{theorem}\label{theoexternalgreedy}
Let $k:X \times X \rightarrow \mathbb{R}\cup\{+\infty\}$ be a
symmetric kernel on a locally compact metric space $X$, $A\subset
X$ be a closed set, and $f:X \rightarrow
\mathbb{R}\cup\{+\infty\}$ be an external field satisfying
$(\ref{condkf})$ in case that $X$ is not compact. Assume that the
Gauss variational problem is solvable and
$\mu\in\mathcal{M}_{f}(A)$ is a solution. Let
$\{\alpha_{N,\mu}^{f}\}$ be a weighted greedy $(f,\mu)$-energy
sequence on $A$. Then
\begin{itemize}
\item[(i)] the following limit
\begin{equation}\label{asympweight}
\lim_{N\rightarrow\infty}\frac{E_{f}(\alpha_{N,\mu}^{f})}{N^{2}}
=V_{f}
\end{equation}
holds.

\item[(ii)] If the equilibrium measure $\mu\in\mathcal{M}_{f}(A)$
is unique, it follows that
\begin{equation}\label{distribweight}
\frac{1}{N}\sum_{a\in\alpha_{N,\mu}^{f}}
\delta_{a}\stackrel{*}{\longrightarrow} \mu\,,\qquad
N\rightarrow\infty\,,
\end{equation}
\begin{equation}\label{eq29}
\lim_{n\rightarrow\infty}\frac{U_{n}^{f}(a_{n})}{n}=V_{f}-\int
f\,d\mu\,,
\end{equation}
where $a_{n}$ is the $n$-th element of the weighted greedy
$(f,\mu)$-energy sequence.
\end{itemize}
\end{theorem}

Conditions (\ref{asympweight})-(\ref{eq29}) are related in the
following way.
\begin{proposition}\label{relationconditions}
Let $k:X \times X \rightarrow \mathbb{R}$ be a real-valued
symmetric kernel on a locally compact metric space $X$, $A\subset
X$ be a closed set, and $f:X \rightarrow
\mathbb{R}\cup\{+\infty\}$ be an external field. Assume that the
Gauss variational problem is solvable and
$\mu\in\mathcal{M}_{f}(A)$ is a solution. Suppose that
$\{b_{n}\}_{n=1}^{\infty}\subset S_{\mu}^{*}$ is a sequence of
points such that
\begin{equation}\label{eq:assumption}
\frac{1}{N}\sum_{n=1}^{N}\delta_{b_{n}}\stackrel{*}{\longrightarrow}
\mu\,,\qquad N\rightarrow\infty\,,
\end{equation}
and set
\begin{equation*}
T_{n}^{f}(x):=\sum_{i=1}^{n-1}k(x,b_{i})+(n-1)f(x)\,,\qquad x\in
A\,,\quad n\geq 2\,.
\end{equation*}
If the following limit
\begin{equation}\label{eq:Tnbn}
\lim_{n\rightarrow\infty}\frac{T_{n}^{f}(b_{n})}{n}=V_{f}-\int
f\,d\mu
\end{equation}
holds, then
\begin{equation}\label{eq:Efb}
\lim_{N\rightarrow\infty}\frac{E_{f}(\{b_{1},\ldots,b_{N}\})}{N^{2}}=V_{f}\,.
\end{equation}
\end{proposition}

\vspace{0.2cm}

Theorem \ref{theoexternalgreedy} can be extended to the following
class of weighted greedy sequences.

\begin{definition}\label{defngreedyprimera2}
Let $m\geq 2$ be a fixed integer. Under the same assumptions of
Definition \ref{defngreedyprimera}, suppose that the Gauss
variational problem is solvable and $\mu\in\mathcal{M}_{f}(A)$ is
an equilibrium measure. A sequence
$(a_{n}=a_{n,m,f,\mu})_{n=1}^{\infty}\subset A$ is called a
\textit{weighted greedy $(m,f,\mu)$-energy sequence} on $A$ if it
is generated inductively in the following way:
\begin{itemize}
\item The first $m$ points $a_{1},\ldots,a_{m}$ are selected so
that $\{a_{1},\ldots,a_{m}\}$ is an optimal weighted $m$-point
configuration on $S_{\mu}^{*}$, i.e.
\begin{equation}\label{defnprimerospuntos}
E_{f}(\{a_{1},\ldots,a_{m}\})\leq E_{f}(\{x_{1},\ldots,x_{m}\})
\end{equation}
for all $(x_{1},\ldots,x_{m})\in S_{\mu}^{*}\times\cdots\times
S_{\mu}^{*}$.

\item Assuming that $a_{1},\ldots,a_{mN}$ have been selected,
where $N\geq 1$ is an integer, the next set of $m$ points
$\{a_{mN+1},\ldots,a_{m(N+1)}\}\subset S_{\mu}^{*}$ are chosen to
minimize the energy functional
\begin{equation}\label{defnprimerospuntos2}
U_{mN}^{(f,m)}(x_{1},\ldots,x_{m}):=\sum_{i=1}^{m}\sum_{l=1}^{mN}k(x_{i},a_{l})
+\sum_{1\leq i<j\leq
m}k(x_{i},x_{j})+((N+1)m-1)\sum_{i=1}^{m}f(x_{i})
\end{equation}
on $S_{\mu}^{*}\times\cdots\times S_{\mu}^{*}$.
\end{itemize}

For every $N\geq 0$, the subindices $mN+1,\ldots,m(N+1)$ are
assigned to the points $a_{mN+1},\ldots,a_{m(N+1)}$ in an
arbitrary order. Let $\alpha_{mN,\mu}^{(f,m)}$ denote the
configuration formed by the first $mN$ points of this sequence.
\end{definition}

In analogy to Definition \ref{defngreedypointsclosed}, we also
introduce the following:
\begin{definition}\label{defngreedypointsclosed2}
Under the same assumptions of Definition
\ref{defngreedypointsclosed}, given an integer $m\geq 2$, a
sequence $(a_{n}=a_{n,m,f})_{n=1}^{\infty}\subset A$ (in case it
exists) is called a \textit{weighted greedy $(m,f)$-energy
sequence} on $A$ if it is obtained inductively as in
$(\ref{defnprimerospuntos})$ and $(\ref{defnprimerospuntos2})$ but
the minimization is taken on $A$. With $\alpha_{mN}^{(f,m)}$ we
denote the configuration $\{a_{1},\ldots,a_{mN}\}$.
\end{definition}

The following result is analogous to Theorem
\ref{theoexternalgreedy}.

\begin{theorem}\label{theogreedygeneral}
Let $m\geq 2$. Under the same assumptions of Theorem
$\ref{theoexternalgreedy}$, assume that
$\{\alpha_{N,\mu}^{(f,m)}\}$ is a weighted greedy
$(m,f,\mu)$-energy sequence on $A$, where
$\mu\in\mathcal{M}_{f}(A)$ is an equilibrium measure solving the
Gauss variational problem. Then
\begin{itemize}
\item[(i)] the following limit
\begin{equation}\label{asympweightgeneral}
\lim_{N\rightarrow\infty}\frac{E_{f}(\alpha_{mN,\mu}^{(f,m)})}{m^2
N^{2}}=V_{f}
\end{equation}
holds.

\item[(ii)] If the equilibrium measure $\mu\in\mathcal{M}_{f}(A)$
is unique, it follows that
\begin{equation}\label{distribweightgeneral}
\frac{1}{mN}\sum_{a\in\alpha_{mN,\mu}^{(f,m)}}
\delta_{a}\stackrel{*}{\longrightarrow} \mu\,,\qquad
N\rightarrow\infty\,,
\end{equation}
\begin{equation}\label{eq29general}
\lim_{N\rightarrow\infty}\frac{U_{mN}^{(f,m)}(a_{mN+1},\ldots,a_{m(N+1)})}{N}=m^{2}(V_{f}-\int
f\,d\mu)\,,
\end{equation}
where $a_{i}$ is the $i$-th element of the weighted greedy
$(m,f,\mu)$-energy sequence.
\end{itemize}
\end{theorem}

\begin{remark}
It is easy to see that $(\ref{distribweightgeneral})$ implies that
\begin{equation*}
\frac{1}{n}\sum_{i=1}^{n}\delta_{a_{i}}\stackrel{*}{\longrightarrow}
\mu\,,\qquad N\rightarrow\infty\,.
\end{equation*}
\end{remark}

Let $p\geq 2$ and consider the Riesz $s$-kernel $k_{s}$ in
$\mathbb{R}^{p}$ (see $(\ref{defnRieszkernel})$) for $s\in(0,p)$.
Assume that $A\subset \mathbb{R}^{p}$ is a closed set and $f$ is
an external field satisfying the following properties:
\begin{equation}\label{conda}
\capp_{s}(\{x\in A: f(x)<+\infty\})>0\,,
\end{equation}
\begin{equation}\label{condb}
\lim_{|x|\rightarrow\infty}f(x)=+\infty\,.
\end{equation}

Using the same arguments employed to prove Theorem I.1.3 in
\cite{SaffTotik} (which concerns the case $p=2$ and $s=0$) and the
fact that $k_{s}$ is positive definite (see \cite[Theorem
1.15]{Landkof}), it is not difficult to see that the Gauss
variational problem on $A$ in the presence of $f$ has a unique
solution $\lambda=\lambda_{s,f}\in\mathcal{M}_{f}(A)$.
Furthermore, the inequality
\begin{equation}\label{Gaussvarineq1Riesz}
U^{\lambda}_{s}(x)+f(x)\leq V_{s,f}-\int f \,d\lambda
\end{equation}
is valid for all $x\in\supp(\lambda)$, where
$V_{s,f}:=I_{s,f}(\lambda)$ denotes the minimal energy constant
$(\ref{Gaussvarprob})$, and
\begin{equation}\label{Gaussvarineq2Riesz}
U^{\lambda}_{s}(x)+f(x)\geq V_{s,f}-\int f \,d\lambda
\end{equation}
holds q.e. on $A$ (relative to the $s$-capacity of sets).

We remark that if $p=2$ and $s=0$ then these properties hold if
$(\ref{condb})$ is replaced by the condition
\begin{equation}\label{condb2}
\lim_{|x|\rightarrow\infty}(f(x)-\log |x|)=+\infty\,.
\end{equation}
The following result holds.

\begin{lemma}\label{lemanBW}
Let $p\geq 2$ and $p-2\leq s< p$. Assume that
$A\subset\mathbb{R}^{p}$ is closed and $f$ satisfies the
conditions $(\ref{conda})$ and $(\ref{condb})$ $($or
$(\ref{condb2})$ in the case $p=2$, $s=0$$)$. Let
$\lambda=\lambda_{s,f}$ be the equilibrium measure solving the
Gauss variational problem on $A$ in the presence of $f$. If
$\{x_{1},\ldots,x_{n}\}\subset \mathbb{R}^{p}$ is an arbitrary
collection of points and
\begin{equation}\label{eq:BWL1}
\sum_{i=1}^{n}\frac{1}{|x-x_{i}|^{s}}+n f(x) \geq M \qquad
\mbox{for}\,\,\mbox{q.e.}\quad x\in \supp(\lambda)\,,
\end{equation}
then for all $x\in \mathbb{R}^{p}$,
\begin{equation}\label{eq:BWL2}
\sum_{i=1}^{n}\frac{1}{|x-x_{i}|^{s}}\geq
M-n(W_{s,f}(\lambda)-U^{\lambda}_{s}(x))\,,
\end{equation}
where $W_{s,f}(\lambda)$ is defined in $(\ref{eq:defnWf})$ and
$U^{\lambda}_{s}$ is the potential associated to $\lambda$.
Moreover, $(\ref{eq:BWL1})$ implies that
\begin{equation}\label{eq:BWL3}
\sum_{i=1}^{n}\frac{1}{|x-x_{i}|^{s}}+n f(x) \geq M \qquad
\mbox{for}\,\,\mbox{q.e.}\quad x\in A\,.
\end{equation}
\end{lemma}

\begin{remark}
The case $p=2$, $s=0$ of Lemma \ref{lemanBW} (the logarithmic
kernel is employed in this case) is known as the
\textit{generalized Bernstein-Walsh lemma} and was proved by H.
Mhaskar and E. Saff in \cite{MhaskarSaff}.
\end{remark}

\begin{corollary}\label{corolario}
Assume that all the assumptions of Lemma $\ref{lemanBW}$ hold. Let
$(a_{n}=a_{n,f})_{n=1}^{\infty}$ be a weighted greedy $f$-energy
sequence on $A$ constructed using the Riesz kernel $k_{s}$ for
$s\in[p-2,p)$. Then this sequence is well-defined and $a_{n}\in
S_{\lambda}^{*}$ for all $n\geq 2$. Moreover, all the asymptotic
properties in Theorem $\ref{theoexternalgreedy}$ are applicable to
this sequence $($replacing $\alpha_{N,\mu}^{f}$ by
$\alpha_{N}^{f}=\{a_{1},\ldots,a_{N}\}$ and $\mu$ by $\lambda$$)$.
\end{corollary}

\begin{corollary}\label{corolario2}
Let $m\geq 2$ and assume that all the assumptions of Lemma
$\ref{lemanBW}$ hold. Let $(a_{n}=a_{n,m,f})_{n=1}^{\infty}$ be a
weighted greedy $(m,f)$-energy sequence on $A$ obtained using the
Riesz kernel $k_{s}$ for $s\in[p-2,p)$. Then this sequence is
well-defined and $a_{n}\in S_{\lambda}^{*}$ for all $n\geq 1$.
Furthermore, all the asymptotic properties in Theorem
$\ref{theogreedygeneral}$ are applicable to this sequence
$($replacing $\alpha_{mN,\mu}^{(f,m)}$ by
$\alpha_{mN}^{(f,m)}=\{a_{1},\ldots,a_{N}\}$ and $\mu$ by
$\lambda$$)$.
\end{corollary}

We remark that the problem of finding an explicit representation
of the solution of a Gauss variational problem in $\mathbb{R}^{p}$
is a difficult task in general. However, there are certain
assumptions on $f$ that could alleviate the difficulty of this
problem, as the following result shows in the case of Newtonian
potentials.

\begin{proposition}\label{propcircweight}
Let $p\geq 3$ and $s=p-2$. Assume that $f$ is a radially symmetric
function $($i.e. $f(x)=f(|x|)$ for all $x\in\mathbb{R}^{p}$$)$
satisfying $(\ref{condb})$. Assume further that, as a function of
$\mathbb{R}_{+}$, $f$ has an absolutely continuous derivative and
obeys one of the following conditions:
\begin{itemize}
\item[(i)] $r^{p-1}f'(r)$ is increasing on $(0,\infty)$;

\item[(ii)] $f$ is convex on $(0,\infty)$.
\end{itemize}
Let $r_{0}$ be the smallest number for which $f'(r)>0$ for all
$r>r_{0}$, and let $R_{0}$ be the smallest solution of
$R_{0}^{p-1}f'(R_{0})=p-2$ $($it is easy to see that $r_{0}<R_{0}$
and $R_{0}$ is finite$)$. If $\lambda_{p-2,f}$ is the solution of
the Gauss variational problem on $A=\mathbb{R}^{p}$ with $f$ as
the external field, then
\begin{equation*}
\supp(\lambda_{p-2,f})=\{x\in\mathbb{R}^{p}:r_{0}\leq |x|\leq
R_{0}\}\,,
\end{equation*}
and $\lambda_{p-2,f}$ is given by
\begin{equation}\label{eq:eqcircweight}
d\lambda_{p-2,f}(x)=\frac{1}{p-2}(r^{p-1}f'(r))'\,dr\,d\sigma_{p-1}(\overline{x})\,,\qquad
x=r\overline{x}\,,\quad r=|x|\,,
\end{equation}
where $d\sigma_{p-1}$ denotes the normalized surface area measure
of the unit sphere $S^{p-1}$ $(\sigma_{p-1}(S^{p-1})=1)$ in
$\mathbb{R}^{p}$. Moreover,
\begin{equation}\label{eq:wflambdacirc}
W_{p-2,f}(\lambda_{p-2,f})=\frac{1}{R_{0}^{p-2}}+f(R_{0})\,,
\end{equation}
and
\begin{equation}\label{eq:extfieldcirc}
U^{\lambda_{p-2,f}}_{p-2}(x)=\left\{
\begin{array}{ccc}
1/R_{0}^{p-2}+f(R_{0})-f(r_{0})\,, & \mbox{if} & |x|\leq r_{0}\,, \\
\\
1/R_{0}^{p-2}+f(R_{0})-f(x)\,, & \mbox{if} & r_{0}< |x| < R_{0}\,, \\
\\
1/|x|^{p-2}\,, & \mbox{if} & |x|\geq R_{0}\,.
\end{array}
\right.
\end{equation}
\end{proposition}

\begin{remark}
The case $p=2$, $s=0$ was analyzed by Mhaskar and Saff in
\cite{MhaskarSaff2} (see Example 3.2 of that paper). See also
\cite{SaffTotik}.
\end{remark}

\section{Numerical experiments}\label{numerical}

\begin{figure}[h]
\begin{minipage}[t]{0.5\linewidth}
\centering
\includegraphics[totalheight=1.5in,keepaspectratio]{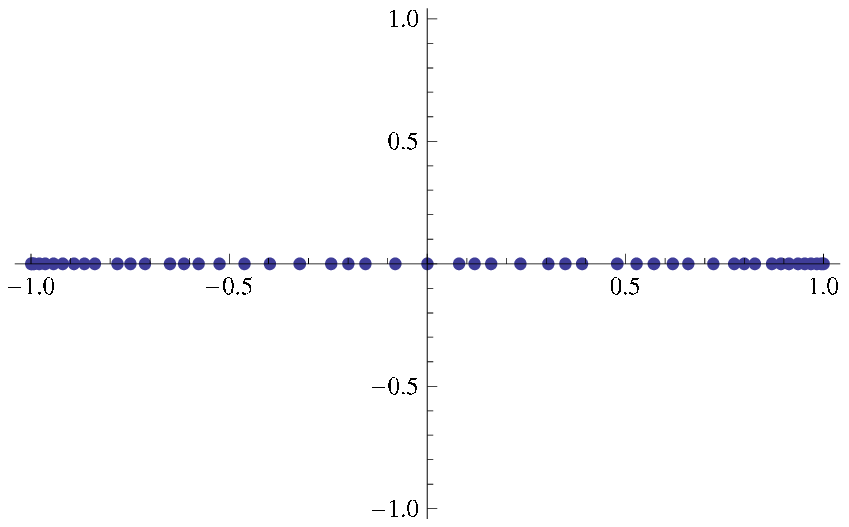}
\caption{$s=0$.} \label{lejapointss0p50}
\end{minipage}%
\begin{minipage}[t]{0.5\linewidth}
\centering
\includegraphics[totalheight=1.5in,keepaspectratio]{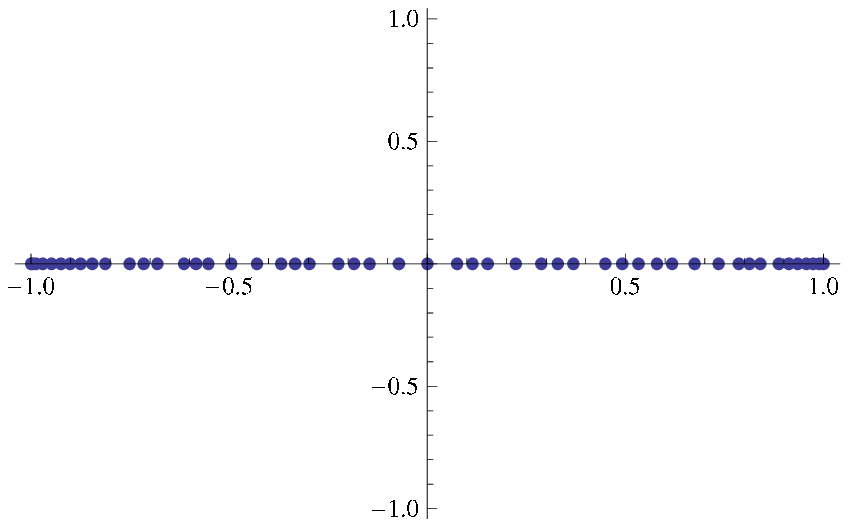}
\caption{$s=0.2$.} \label{lejapointss02p50}
\end{minipage}
\end{figure}

\begin{figure}[h]
\begin{minipage}[t]{0.5\linewidth}
\centering
\includegraphics[totalheight=1.5in,keepaspectratio]{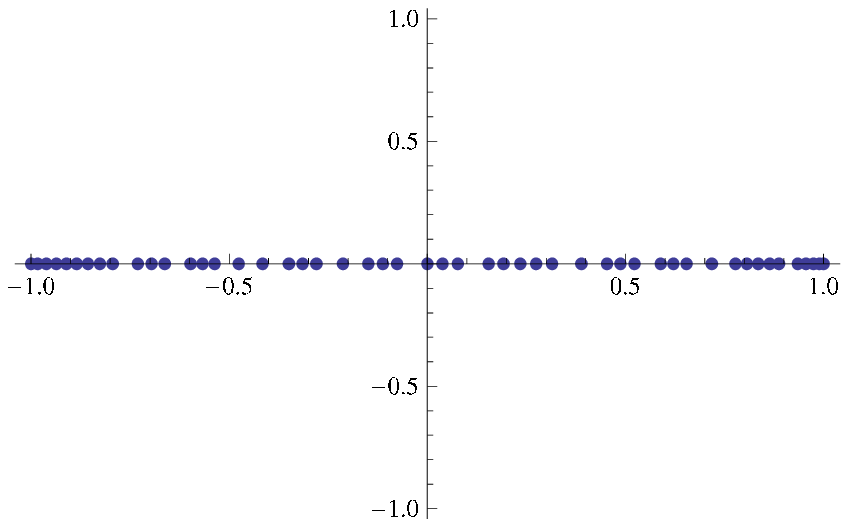}
\caption{$s=0.4$.} \label{lejapointss04p50}
\end{minipage}%
\begin{minipage}[t]{0.5\linewidth}
\centering
\includegraphics[totalheight=1.5in,keepaspectratio]{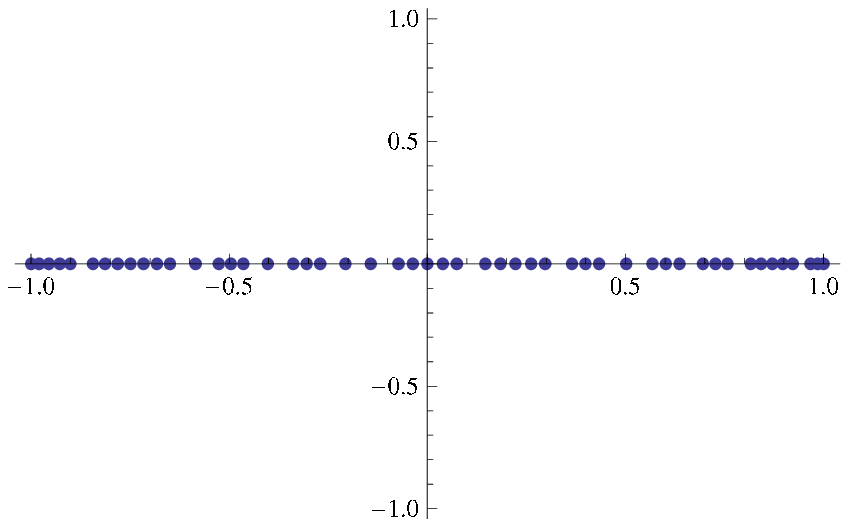}
\caption{$s=0.6$.} \label{lejapointss06p50}
\end{minipage}
\end{figure}

\begin{figure}[h]
\centering
\includegraphics[totalheight=1.5in,keepaspectratio]{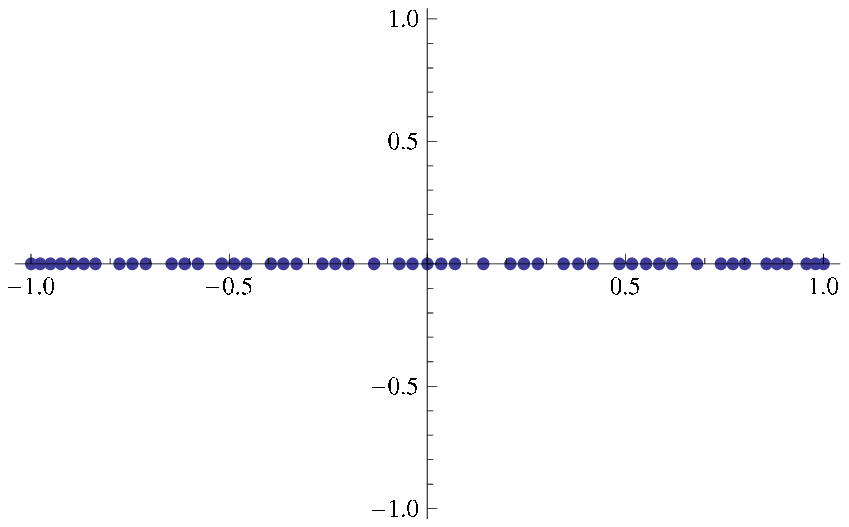}
\caption{$s=0.8$.} \label{lejapointss08p50}
\end{figure}

We first present in Figures
$\ref{lejapointss0p50}$--$\ref{lejapointss08p50}$ some plots of
greedy energy points on $A=[-1,1]$, constructed under the
following conditions: $m=1$, $k=k_{s}$ (the Riesz $s$-kernel), and
$f\equiv 0$. In all these examples the total number of points is
$50$, and the initial point is selected to be $a_{1}=-1$. The
values of $s$ are indicated. Under the above conditions, the
equilibrium measure is given by
\[
d\lambda_{s}(x)=\frac{\Gamma\big(1+\frac{s}{2}\big)}
{\sqrt{\pi}\,\Gamma\big(\frac{1+s}{2}\big)}\,(1-x^2)^{(s-1)/2}dx,\qquad
-1\leq x\leq 1,\quad s\in[0,1),
\]
(cf. \cite{Landkof}).


We now present some plots of weighted greedy energy points. The
following example shows the first $50$ points of a weighted greedy
$f$-energy sequence on $A=[-1,1]$ (see Definition
\ref{defngreedypointsclosed}) for the logarithmic kernel $k_{0}$
and the external field
\begin{equation}\label{externalfield1}
f(x)=|x|,\qquad x\in[-1,1].
\end{equation}
The first point selected for this sequence was again $a_{1}=-1$.
Observe that the points are much more numerous near the origin,
since $f$ takes the lowest value there.

\begin{figure}[h!]
\begin{center}
\includegraphics[totalheight=1.5in,keepaspectratio]{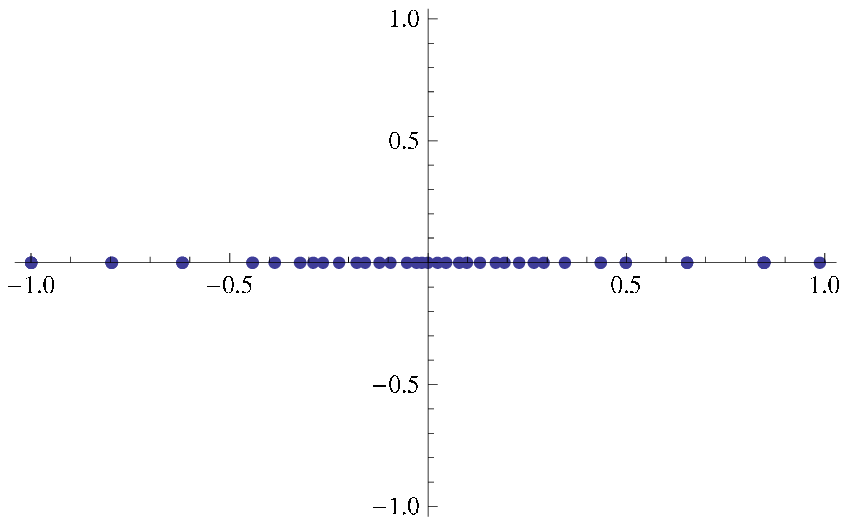}
\caption{$50$ weighted greedy $f$-energy points for the
logarithmic kernel and the external field
$(\ref{externalfield1})$.} \label{lejalogAbs50}
\end{center}
\end{figure}

The next two examples are also weighted greedy $f$-energy
sequences on $A=[-1,1]$ for the logarithmic kernel $k_{0}$, but
now the external field is
\begin{equation}\label{externalfield2}
f(x)=-\log(w(x)),\qquad
w(x)=(1-x)^{\lambda_{1}}(1+x)^{\lambda_{2}},\qquad
\lambda_{1},\lambda_{2}>0.
\end{equation}
It is known (cf. \cite[page 241]{SaffTotik}) that in this case the
equilibrium measure is
\[
d\mu_{\lambda_1,\lambda_2}(x)=\frac{1}{\pi}\frac{(1+\lambda_{1}
+\lambda_{2})}{1-x^2} \sqrt{(x-a)(b-x)},\qquad a\leq x\leq b,
\]
with support $\supp(\mu_{\lambda_{1},\lambda_{2}})=[a,b]$, where
\[
a=\theta_{2}^2-\theta_{1}^2-\sqrt{\Delta},\qquad
b=\theta_{2}^2-\theta_{1}^{2}+\sqrt{\Delta},
\]
and
\[
\theta_{1}:=\frac{\lambda_{1}}{1+\lambda_{1}+\lambda_{2}},\quad
\theta_{2}:=\frac{\lambda_{2}}{1+\lambda_{1}+\lambda_{2}},\quad
\Delta:=[1-(\theta_1+\theta_2)^2][1-(\theta_1-\theta_2)^2].
\]
The following example corresponds to the choice $\lambda_{1}=2,
\lambda_{2}=1$. The point $a_{1}$ is the origin, and the number of
points shown is again $50$. In this case, $a\approx-0.83$ and
$b\approx 0.45$.

\begin{figure}[h]
\begin{center}
\includegraphics[totalheight=1.5in,keepaspectratio]{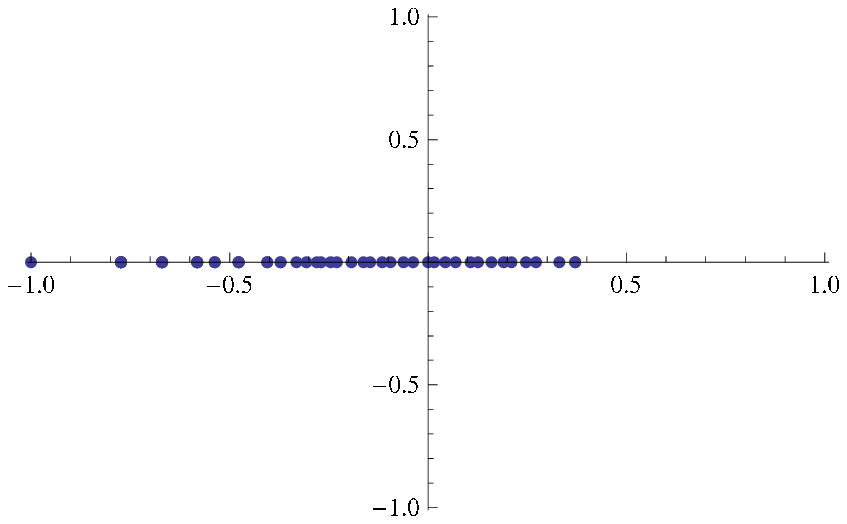}
\caption{$50$ weighted greedy $f$-energy points for the
logarithmic kernel and the external field $(\ref{externalfield2})$
with parameters $\lambda_{1}=2, \lambda_{2}=1$.}
\label{Jacobi2150nuevo}
\end{center}
\end{figure}

In the following example we choose $\lambda_{1}=4, \lambda_{2}=1$,
and again $a_{1}=0$.

\begin{figure}[h!]
\begin{center}
\includegraphics[totalheight=1.5in,keepaspectratio]{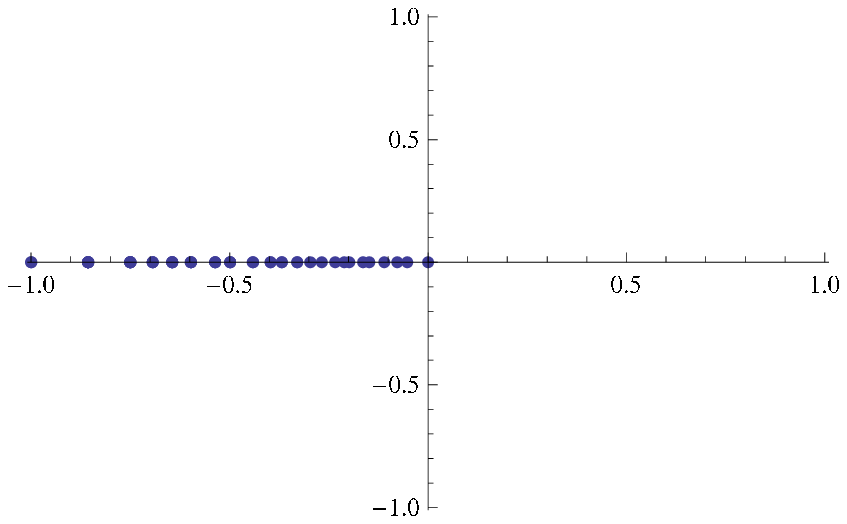}
\caption{$50$ weighted greedy $f$-energy points for the
logarithmic kernel and the external field $(\ref{externalfield2})$
with parameters $\lambda_{1}=4, \lambda_{2}=1$.}
\label{Jacobi4150otro}
\end{center}
\end{figure}

Observe that now all the points were pushed to the interval
$[-1,0]$! Another interesting phenomenon can be observed, which is
that many points are almost coincident. In this example, $a\approx
-0.89$ and $b\approx 0.062$.

\begin{figure}[h]
\begin{center}
\includegraphics[totalheight=2.5in,keepaspectratio]{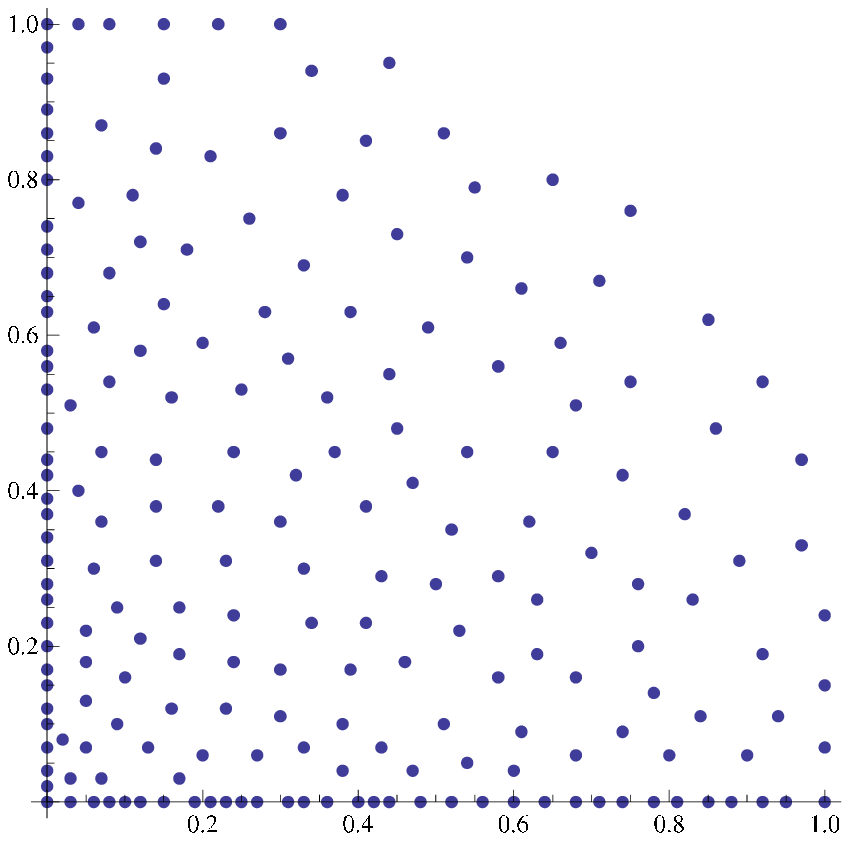}
\caption{$200$ weighted greedy $f$-energy points on $[0,1]^2$ for
$s=0.8$ and the external field $f(x,y)=x^2+y^2$.}
\label{puntoslejaweightedsquare}
\end{center}
\end{figure}

In Figure \ref{puntoslejaweightedsquare} we show the first $200$
points of a weighted greedy $f$-energy sequence on $A=[0,1]^2$.
The initial point is the origin, $s=0.8$ and the external field is
$f(x,y)=x^2+y^2$, $(x,y)\in A$.

\section{Proofs}\label{Proofs}
\noindent{\textsc{Proof of Theorem \ref{primer}.}} Our first goal
is to show that
\begin{equation}\label{eq25}
\limsup_{N\rightarrow\infty}\frac{E_{f}(\omega_{N}^{*})}{N^{2}}\leq
V_{f}\,.
\end{equation}
Let $\nu\in\mathcal{M}_{f}(A)$ be arbitrary, and consider the
measure $\lambda:=\bigotimes_{j=1}^{N}\nu$ on the pro\-duct space
$X^{N}$. Define the function
$h:X^{N}\rightarrow\mathbb{R}\cup\{+\infty\}$ by
$h(x_{1},\ldots,x_{N}):=E_{f}(\{x_{1},\ldots,x_{N}\})$. Therefore,
$E_{f}(\omega_{N}^{*})\leq h(x_{1},\ldots,x_{N})$ for all
$(x_{1},\ldots,x_{N})\in A^{N}$. Integrating with respect to
$\lambda$ it follows that
\begin{equation*}
E_{f}(\omega_{N}^{*})\leq \int_{A^{N}}
h(x_{1},\ldots,x_{N})\,d\lambda(x_{1},\ldots,x_{N})
\end{equation*}
\begin{equation*}
=\int_{A^{N}}\sum_{1\leq i\neq j\leq
N}k(x_{i},x_{j})\,d\lambda(x_{1},\ldots,x_{N})
+2(N-1)\int_{A^{N}}\sum_{i=1}^{N}f(x_{i})\,d\lambda(x_{1},\ldots,x_{N})
\end{equation*}
\begin{equation*}
=\sum_{1\leq i\neq j\leq
N}\int_{A^{2}}k(x_{i},x_{j})\,d\nu(x_{i})\,d\nu(x_{j})+2(N-1)\sum_{i=1}^{N}
\int_{A}f(x_{i})\,d\nu(x_{i})
\end{equation*}
\begin{equation*}
=N(N-1)\Big(\int_{A^{2}}k(x,y)\,d\nu(x)\,d\nu(y)+2\int_{A}f(x)\,d\nu(x)\Big)
=N(N-1)I_{f}(\nu)\,.
\end{equation*}
Taking the infimum over $\nu\in\mathcal{M}_{f}(A)$ we obtain that
$E_{f}(\omega_{N}^{*})\leq N(N-1)V_{f}$, and therefore
(\ref{eq25}) holds.

Next we show that
\begin{equation}\label{eq26}
V_{f}\leq
\liminf_{N\rightarrow\infty}\frac{E_{f}(\omega_{N}^{*})}{N^{2}}
\end{equation}
and at the same time we verify (\ref{distribweightmin}). Let
$\omega_{N}^{*}=\{x_{1},\ldots,x_{N}\}$ and define
\begin{equation*}
\nu_{N}:=\frac{1}{N}\sum_{i=1}^{N}\delta_{x_{i}}\,.
\end{equation*}
Assume that $g_{n}:A\times A\rightarrow\mathbb{R}$ is a sequence
of non-decreasing continuous functions that converges pointwise to
$k$ on $A$. We fix $n$. Then
\begin{equation}\label{eqfinal}
\int\int g_{n}(x,y)\,d\nu_{N}(x)\,d\nu_{N}(y)+2\int f\, d\nu_{N}
\end{equation}
\begin{equation*}
=\frac{1}{N^{2}}\Big(\sum_{i=1}^{N}g_{n}(x_{i},x_{i})+\sum_{1\leq
i\neq j\leq N}g_{n}(x_{i},x_{j})+2N\sum_{i=1}^{N}f(x_{i})\Big)
\end{equation*}
\begin{equation*}
\leq\frac{1}{N^{2}}\Big(\sum_{i=1}^{N}(g_{n}(x_{i},x_{i})+2f(x_{i}))+\sum_{1\leq
i\neq j\leq N}k(x_{i},x_{j})+2(N-1)\sum_{i=1}^{N}f(x_{i})\Big)
\end{equation*}
\begin{equation*}
=\frac{1}{N^{2}}\Big(\sum_{i=1}^{N}(g_{n}(x_{i},x_{i})+2f(x_{i}))
+E_{f}(\omega_{N}^{*})\Big)\,. \end{equation*} Let
$C:=\inf\{k(x,y):(x,y)\in A^{2}\}$ and $D:=\inf\{f(x):x\in A\}$.
Both $C$ and $D$ are finite since $A$ is compact and $k$ and $f$
are lower semicontinuous. Using $E_{f}(\omega_{N}^{*})\leq
N(N-1)V_{f}$ we obtain
\begin{equation}\label{eq:estimatesf}
ND\leq \sum_{i=1}^{N}f(x_{i})\leq \frac{N}{2}(V_{f}-C)\,.
\end{equation}
By the compactness of $A$ and the continuity of $g_{n}$, there
exists a constant $M_{n}>0$ such that
\begin{equation*}
\sum_{i=1}^{N}|g_{n}(x_{i},x_{i})|\leq N\,M_{n}\,.
\end{equation*}
In particular,
\begin{equation}\label{asimpdelosg}
\frac{\sum_{i=1}^{N}g_{n}(x_{i},x_{i})}{N^{2}}\longrightarrow
0,\qquad N \longrightarrow\infty\,.
\end{equation}
From (\ref{eq:estimatesf}) and (\ref{asimpdelosg}) we conclude
that
\begin{equation}\label{eq27}
\frac{\sum_{i=1}^{N}(g_{n}(x_{i},x_{i})+2f(x_{i}))}{N^{2}}\longrightarrow
0,\qquad N \longrightarrow\infty\,.
\end{equation}

Let $\nu\in\mathcal{M}_{1}(A)$ be a cluster point of the sequence
$\{\nu_{N}\}$ in the weak-star topology. Then there exists a
subsequence $\{\nu_{N}\}_{N\in\mathcal{N}}$ that converges
weak-star to $\nu$ (cf. \cite[Lemma 1.2.1]{Fuglede}). Therefore
\begin{equation}\label{eq:estimatesintgn}
\int\int g_{n}(x,y)\,d\nu(x)\,d\nu(y)+2\int f(x)\,d\nu(x)
\end{equation}
\begin{equation*}
\leq\liminf_{N\in\mathcal{N}}\Big(\int\int
g_{n}(x,y)\,d\nu_{N}(x)\,d\nu_{N}(y)+2\int
f(x)\,d\nu_{N}(x)\Big)\,. \end{equation*} Now we apply
(\ref{eq:estimatesintgn}), (\ref{eqfinal}), (\ref{eq27}) and
(\ref{eq25}) to obtain
\begin{equation*}
\int\int g_{n}(x,y)\,d\nu(x)\,d\nu(y)+2\int f(x)\,d\nu(x)\leq
V_{f}\,. \end{equation*} From the monotone convergence theorem we
conclude that
\begin{equation*}
I_{f}(\nu)=\lim_{n\rightarrow\infty}\int\int
g_{n}(x,y)\,d\nu(x)\,d\nu(y)+2\int f(x)\,d\nu(x)\leq V_{f}\,.
\end{equation*} Therefore $\nu=\mu$, the equilibrium measure. Since
$\mu$ is the only cluster point of $\{\nu_{N}\}$,
(\ref{distribweightmin}) follows.

Using (\ref{eqfinal}) we have
\begin{equation*}
\int\int g_{n}(x,y)\,d\mu(x)\,d\mu(y)+2\int f(x)\,d\mu(x)
\end{equation*}
\begin{equation*}
\leq \liminf_{N\rightarrow\infty}\frac{1}{N^{2}}
\Big(\sum_{i=1}^{N}(g_{n}(x_{i},x_{i})+2f(x_{i}))+E_{f}(\omega_{N}^{*})
\Big)
=\liminf_{N\rightarrow\infty}\frac{1}{N^{2}}E_{f}(\omega_{N}^{*})\,,
\end{equation*} from which (\ref{eq26}) follows. Finally,
(\ref{asympweightmin}) is a consequence of (\ref{eq26}) and
(\ref{eq25}).\hfill $\Box$

\begin{lemma}\label{lemconv}
Let $k:X \times X \rightarrow \mathbb{R}\cup\{+\infty\}$ be a
symmetric kernel on a locally compact metric space $X$, $A\subset
X$ be a compact set, and $f:X \rightarrow
\mathbb{R}\cup\{+\infty\}$ be an external field. Assume that the
Gauss variational problem is solvable and
$\mu\in\mathcal{M}_{f}(A)$ is a solution. Let
$\{\tau_{n}\}\subset\mathcal{M}_{1}(S_{\mu}^{*})$ be a sequence of
measures that converges to $\mu$ in the weak-star topology. Then
\begin{equation}\label{eqlemconv}
\lim_{n\rightarrow\infty}\int f\,d\tau_{n}=\int f\,d\mu\,.
\end{equation}
\end{lemma}
\begin{proof} Since $f$ and $U^{\mu}$ are lower semicontinuous we have
\begin{equation*}
\int f\,d\mu \leq \liminf_{n\rightarrow\infty}\int f\,d\tau_{n}\,,
\end{equation*}
\begin{equation*}
\limsup_{n\rightarrow\infty}\int(W_{f}(\mu)-U^{\mu})\,d\tau_{n}
\leq\int(W_{f}(\mu)-U^{\mu})\,d\mu\,. \end{equation*} In addition,
for $x\in S_{\mu}^{*}$ the inequality $f(x)\leq
W_{f}(\mu)-U^{\mu}(x)$ holds, and therefore
\begin{equation*}
\limsup_{n\rightarrow\infty}\int f\,d\tau_{n}\leq
\limsup_{n\rightarrow\infty}\int(W_{f}(\mu)-U^{\mu})\,d\tau_{n}\,.
\end{equation*} By (\ref{Gaussvarineq1}) and (\ref{Gaussvarineq2}),
$f=W_{f}(\mu)-U^{\mu}$ q.e. on $S_{\mu}$, and since $\mu$ has
finite energy this equality holds $\mu$-a.e. Thus
\begin{equation*}
\int f\,d\mu=\int(W_{f}(\mu)-U^{\mu})\,d\mu\,, \end{equation*} and
(\ref{eqlemconv}) follows.
\end{proof}

\noindent{\textsc{Proof of Theorem \ref{theoexternalgreedy}.}} To
prove this result we follow closely ideas from chapter V of
\cite{SaffTotik}. By definition,
\begin{equation*}
U_{n}^{f}(a_{n})\leq U_{n}^{f}(x) \,\qquad\mbox{for all}\quad x\in
S_{\mu}^{*}\,,\quad n\geq 2\,.\end{equation*} We have, for any
$x\in S_{\mu}^{*}$,
\begin{equation*}
E_{f}(\alpha_{N,\mu}^{f})=2\sum_{1\leq i<j\leq
N}k(a_{i},a_{j})+2(N-1)\sum_{i=1}^{N}f(a_{i})
\end{equation*}
\begin{equation*}
=2\sum_{j=2}^{N}\Big(\sum_{i=1}^{j-1}k(a_{i},a_{j})+(j-1)f(a_{j})
+\sum_{i=1}^{j-1}f(a_{i}) \Big)\,
\end{equation*}
\begin{equation*}
=2\sum_{j=2}^{N}\Big(U_{j}^{f}(a_{j})+\sum_{i=1}^{j-1}f(a_{i})\Big)
\leq
2\sum_{j=2}^{N}\Big(U_{j}^{f}(x)+\sum_{i=1}^{j-1}f(a_{i})\Big)
\end{equation*}
\begin{equation*}
=2\sum_{j=2}^{N}\sum_{i=1}^{j-1}\Big(k(x,a_{i})+f(x)+f(a_{i})\Big)\,.
\end{equation*}
We now integrate with respect to $\mu$ to obtain
\begin{equation*}
E_{f}(\alpha_{N,\mu}^{f})\leq
2\sum_{j=2}^{N}\sum_{i=1}^{j-1}\Big(U^{\mu}(a_{i})+\int f
\,d\mu+f(a_{i})\Big)\,.
\end{equation*}
Taking into account that $U^{\mu}(a_{i})+f(a_{i})\leq W_{f}(\mu)$
for all $i$ and $W_{f}(\mu)+\int f \,d\mu= V_{f}$, it follows that
\begin{equation}\label{eqmitad}
E_{f}(\alpha_{N,\mu}^{f})\leq N(N-1)V_{f}\,.
\end{equation}
Now, if $\{\omega_{N}^{*}\}$ is a sequence of optimal weighted
$N$-point configurations on $S_{\mu}^{*}$, then
$E_{f}(\omega_{N}^{*})\leq E_{f}(\alpha_{N,\mu}^{f})$ for all $N$.
Therefore (\ref{asympweight}) is a consequence of (\ref{eqmitad})
and (\ref{asympweightmin}).

Throughout the rest of the proof we assume that the equilibrium
measure $\mu\in\mathcal{M}_{f}(A)$ is unique. Consider the
sequence of normalized counting measures
\begin{equation*}
\nu_{N}:=\frac{1}{N}\sum_{a\in\alpha_{N,\mu}^{f}}\delta_{a}\,.
\end{equation*}
As in the proof of Theorem \ref{primer}, we select a sequence
$g_{n}:S_{\mu}^{*}\times S_{\mu}^{*}\rightarrow\mathbb{R}$ of
non-decreasing continuous functions that converges pointwise to
$k$ on $S_{\mu}^{*}$. We have, as in (\ref{eqfinal}),
\begin{equation*}
\int\int g_{n}(x,y)\,d\nu_{N}(x)\,d\nu_{N}(y)+
2\,\frac{N-1}{N}\int f \,d\nu_{N}\leq
\frac{\sum_{i=1}^{N}g_{n}(a_{i},a_{i})
+E_{f}(\alpha_{N,\mu}^{f})}{N^2}\,.\end{equation*} Let
$\{\nu_{N}\}_{N\in\mathcal{N}}$ be a subsequence that converges in
the weak-star topology to a measure
$\lambda\in\mathcal{M}_{1}(A)$. By the lower-semicontinuity of
$f$,
\begin{equation*}
\int f \,d \lambda \leq \liminf_{N\in\mathcal{N}}\int f
\,d\nu_{N}\,.\end{equation*} Thus from (\ref{asimpdelosg}) and
(\ref{asympweight}) we conclude that
\begin{equation*}
\int\int g_{n}(x,y)\,d\lambda(x)\,d\lambda(y)+2\,\int f \,d\lambda
\leq V_{f}\,.\end{equation*} Now we let $n\rightarrow\infty$ and
obtain
\begin{equation*}
I_{f}(\lambda)=\int\int k(x,y)\,d\lambda(x)\,d\lambda(y)+2\,\int f
\,d\lambda \leq V_{f}\,.\end{equation*} It follows that
$\lambda\in\mathcal{M}_{f}(A)$ and $\lambda$ is an equilibrium
measure. By hypothesis there is only one equilibrium measure, thus
$\lambda=\mu$ and (\ref{distribweight}) is proved.

We next show (\ref{eq29}). First,
\begin{equation}\label{eq30}
\sum_{i=2}^{N}U_{i}^{f}(a_{i})=
\frac{1}{2}\,E_{f}(\alpha_{N,\mu}^{f})-\sum_{i=1}^{N}(N-i)f(a_{i})\,.
\end{equation}
By (\ref{distribweight}) and Lemma \ref{lemconv},
\begin{equation}\label{eq:asympmeanf}
\lim_{N\rightarrow\infty}\frac{1}{N}\sum_{i=1}^{N}f(a_{i})=\int
f\,d\mu\,.
\end{equation}
This implies that
\begin{equation}\label{eq31}
\lim_{N\rightarrow\infty}\frac{2}{(N-1)N}\sum_{i=1}^{N}(N-i)f(a_{i})
=\int f\,d\mu\,.
\end{equation}
Applying (\ref{asympweight}), (\ref{eq30}), and (\ref{eq31}), we
obtain
\begin{equation}\label{eq32}
\lim_{N\rightarrow\infty}\frac{2}{(N-1)N}\sum_{i=2}^{N}
U_{i}^{f}(a_{i})=V_{f}-\int f\,d\mu=W_{f}(\mu)\,.
\end{equation}
For every $n\geq 1$,
\begin{equation*}
\frac{U_{n+1}^{f}(a_{n+1})}{n}=\inf_{x\in
S_{\mu}^{*}}\Big\{\frac{1}{n}\,\sum_{i=1}^{n}k(x,a_{i})+f(x)\Big\}\,.
\end{equation*}
Integrating this expression with respect to $\mu$ it follows that
\begin{equation}\label{eq34}
\frac{U_{n+1}^{f}(a_{n+1})}{n}
\leq\frac{1}{n}\,\sum_{i=1}^{n}U^{\mu}(a_{i})+\int f\,d\mu\leq
W_{f}(\mu)+\int f\,d\mu-\frac{1}{n}\sum_{i=1}^{n}f(a_{i})\,.
\end{equation}
Let
\begin{equation*}
\rho_{n}:=\int f\,d\mu-\frac{1}{n}\sum_{i=1}^{n}f(a_{i})\,,\qquad
n\geq 1\,.
\end{equation*}
On the other hand, for every $n\geq 2$,
\begin{equation}\label{eq35}
U_{n+1}^{f}(a_{n+1})\geq U_{n}^{f}(a_{n})+L\,,
\end{equation}
where $L:=\inf\{k(x,a)+f(x):a,x\in S_{\mu}^{*}\}$. We may assume
that $L\leq -1$.

Let $\epsilon\in(0,1)$. Assume that $m$ is an integer such that
\begin{equation}\label{eqcondition}
\frac{U_{m+1}^{f}(a_{m+1})}{m}<W_{f}(\mu)-\epsilon\,.
\end{equation}
Applying (\ref{eq35}) repeatedly we obtain for
$(1+\epsilon/(3L))m\leq i\leq m$,
\begin{equation*}
\frac{U_{i+1}^{f}(a_{i+1})}{m}\leq
W_{f}(\mu)-\epsilon-\frac{(m-i)L}{m}\leq
W_{f}(\mu)-\epsilon+\frac{\epsilon/3}{1+\epsilon/(3L)} \leq
W_{f}(\mu)-\frac{\epsilon}{2}\,,
\end{equation*}
and so
\begin{equation*}
\frac{U_{i+1}^{f}(a_{i+1})}{i}\leq
\frac{m}{i}(W_{f}(\mu)-\epsilon/2)\leq
\frac{m}{i}W_{f}(\mu)-\frac{\epsilon}{2}\,.
\end{equation*}
Taking into account (\ref{eq34}) and the previous inequality,
\begin{equation}\label{eq:imp1}
\frac{2}{(m+1)\,m}\sum_{i=1}^{m}U_{i+1}^{f}(a_{i+1})\leq
\frac{2}{(m+1)\,m}\sum_{1\leq
i<(1+\epsilon/(3L))m}i(W_{f}(\mu)+\rho_{i})
\end{equation}
\begin{equation*}
+\frac{2}{(m+1)\,m}\sum_{(1+\epsilon/(3L))m\leq i\leq m}m\,
W_{f}(\mu)-\frac{\epsilon}{2}\frac{2}{(m+1)\,m}\sum_{(1+\epsilon/(3L))m\leq
i\leq m}i\,.
\end{equation*}
Furthermore, it is easy to see that
\begin{equation}\label{eq:imp2}
-\frac{\epsilon}{2}\frac{2}{(m+1)\,m}\sum_{(1+\epsilon/(3L))m\leq
i\leq
m}i\leq\frac{\epsilon^2}{6L(m+1)}\Big(1+2\,m+\frac{m\epsilon}{3L}\Big)
\end{equation}
\begin{equation*}
\leq \frac{\epsilon^{2}(1+\epsilon/(3L))}{6L}\,.
\end{equation*}

By $(\ref{eq:asympmeanf})$ we know that $\rho_{n}\longrightarrow
0$ as $n\longrightarrow\infty$, which implies that
\begin{equation}\label{eq:asymprho}
\lim_{N\rightarrow\infty}\frac{2}{(N+1)\,N}\sum_{1\leq
i<(1+\epsilon/(3L))N}i\,\rho_{i}=0\,.
\end{equation}
If $W_{f}(\mu)\leq 0$, then
\begin{equation*}
\frac{2}{(m+1)\,m}\Big\{\sum_{1\leq
i<(1+\epsilon/(3L))m}i\,W_{f}(\mu)+\sum_{(1+\epsilon/(3L))m\leq
i\leq m}m \,W_{f}(\mu) \Big\}\leq W_{f}(\mu)\,,
\end{equation*}
and hence it follows from (\ref{eq:imp1}) and (\ref{eq:imp2}) that
\begin{equation}\label{eq:imp3}
\frac{2}{m(m+1)}\sum_{i=1}^{m}U_{i+1}^{f}(a_{i+1})
\end{equation}
\begin{equation*}
\leq
W_{f}(\mu)+\frac{\epsilon^{2}(1+3\epsilon/(3L))}{6L}+\frac{2}{(m+1)\,m}\sum_{1\leq
i<(1+\epsilon/(3L))m}i\,\rho_{i}\,.
\end{equation*}
Since the second term of the right-hand side of (\ref{eq:imp3}) is
a negative constant, applying (\ref{eq:asymprho}), (\ref{eq32}),
and (\ref{eq:imp3}), it follows that there are finitely many
integers $m$ satisfying (\ref{eqcondition}). This together with
(\ref{eq34}) implies (\ref{eq29}).

Now we assume that $W_{f}(\mu)>0$. It is easy to verify that
\begin{equation*}
\frac{2}{(m+1)m}\Big\{\sum_{1\leq
i<(1+\epsilon/(3L))m}i\,W_{f}(\mu)+\sum_{(1+\epsilon/(3L))m\leq
i\leq m}m\, W_{f}(\mu) \Big\}
\end{equation*}
\begin{equation*}
\leq \Big(1+\frac{2}{m+1}+\frac{\epsilon}{3L(m+1)}
+\frac{\epsilon^{2}m}{9(m+1)L^{2}}\Big)W_{f}(\mu)\,,
\end{equation*}
and so, from (\ref{eq:imp1}) and (\ref{eq:imp2}), we deduce that
\begin{equation*}
\frac{2}{(m+1)\,m}\sum_{i=1}^{m}U_{i+1}^{f}(a_{i+1})\leq
\Big(1+\frac{2}{m+1}+\frac{\epsilon}{3L(m+1)}+\frac{\epsilon^{2}m}{9(m+1)L^{2}}\Big)W_{f}(\mu)
\end{equation*}
\begin{equation*}
+\frac{\epsilon^{2}(1+\epsilon/(3L))}{6L}+\frac{2}{(m+1)\,m}\sum_{1\leq
i<(1+\epsilon/(3L))m}i\,\rho_{i}\,.
\end{equation*}
If we assume that there is an infinite sequence $\mathcal{N}$ of
integers $m$ satisfying (\ref{eqcondition}), applying the last
inequality and (\ref{eq:asymprho}), we obtain
\begin{equation}\label{eq:ocho}
\limsup_{m\in\mathcal{N}}\frac{2}{(m+1)\,m}\sum_{i=1}^{m}U_{i+1}^{f}(a_{i+1})
\leq W_{f}(\mu)+\frac{\epsilon^{2}\,W_{f}(\mu)}
{9\,L^{2}}+\frac{\epsilon^{2}(1+\epsilon/(3L))}{6L}\,.
\end{equation}
We may assume without loss of generality that
$L<-(1+2\,W_{f}(\mu))/3$. Then the right-hand side of
(\ref{eq:ocho}) is a constant strictly less than $W_{f}(\mu)$,
which contradicts (\ref{eq32}). This concludes the proof of
(\ref{eq29}). \hfill $\Box$

\vspace{0.2cm}

\noindent{\textsc{Proof of Proposition \ref{relationconditions}.}}
We know that
\begin{equation}\label{primero}
E_{f}(\{b_{1},\ldots,b_{N}\})=2\sum_{i=2}^{N}T_{i}^{f}(b_{i})+2\sum_{i=1}^{N}(N-i)f(b_{i})\,.
\end{equation}
Since $k$ is real-valued and $\{b_{n}\}_{n=1}^{\infty}\subset
S_{\mu}^{*}$, we have that $T_{n}^{f}(b_{n})<+\infty$ for all $n$.
From (\ref{eq:Tnbn}) it follows that
\begin{equation}\label{segundo}
\lim_{N\rightarrow\infty}\frac{2}{N(N-1)}\sum_{i=2}^{N}
T_{i}^{f}(b_{i})=V_{f}-\int f\,d\mu\,,
\end{equation}
and applying (\ref{eq:assumption}) and Lemma \ref{lemconv}, we
obtain
\begin{equation}\label{tercero}
\lim_{N\rightarrow\infty}\frac{2}{N(N-1)}\sum_{i=1}^{N}(N-i)\,f(b_{i})=\int
f\,d\mu\,.
\end{equation}
Therefore (\ref{eq:Efb}) is a consequence of
(\ref{primero})-(\ref{tercero}).\hfill $\Box$

\vspace{0.2cm}

The proof of Theorem \ref{theogreedygeneral} is similar to that of
Theorem \ref{theoexternalgreedy}, and consequently we only give a
sketch of its proof. The details are left to the reader.

\vspace{0.2cm}

\noindent{\textsc{Sketch of the proof of Theorem
\ref{theogreedygeneral}}.} In order to prove
$(\ref{asympweightgeneral})$, we use the fact that
\begin{equation*}
E_{f}(\alpha_{mN,\mu}^{(f,m)})=2\sum_{j=1}^{N-1}\Big[U_{jm}^{(f,m)}(a_{jm+1},\ldots,a_{(j+1)m})
+m\sum_{r=1}^{j-1}\sum_{l=1}^{m}f(a_{rm+l})\Big]+\phi_{m,N}\,,
\end{equation*}
where
\begin{equation*}
\phi_{m,N}=E(\{a_{1},\ldots,a_{m}\})+2(mN-1)\sum_{i=1}^{m}f(a_{i})\,.
\end{equation*}
Using the definition of $\{a_{jm+1},\ldots,a_{(j+1)m}\}$ and
integrating the resulting inequality by
$d\mu(x_{m+1})\times\cdots\times d\mu(x_{mN})$ it follows that
\begin{equation*}
E_{f}(\alpha_{mN,\mu}^{(f,m)})\leq
m^{2}(N-1)(N-2)W_{f}(\mu)+m^{2}(N+1)N\int f\,d\mu+o(N^{2})\,.
\end{equation*}
This inequality and $(\ref{asympweightmin})$ imply
$(\ref{asympweightgeneral})$. The asymptotic expression
(\ref{distribweightgeneral}) is an application of
(\ref{asympweightgeneral}).

To prove $(\ref{eq29general})$ we use the inequalities
\begin{equation*}
\frac{U_{mN}^{(f,m)}(a_{mN+1},\ldots,a_{m(N+1)})}{mN}\leq
m\,W_{f}(\mu)+\rho_{m,N}\,,
\end{equation*}
\begin{equation*}
U_{m(N+1)}^{(f,m)}(a_{m(N+1)+1},\ldots,a_{m(N+2)})\geq U_{m
N}^{(f,m)}(a_{m N+1},\ldots,a_{m(N+1)})+m^2 L\,,
\end{equation*}
where
\[
\rho_{m,N}=m\Big(\int
f\,d\mu-\frac{1}{mN}\sum_{i=1}^{mN}f(a_{i})\Big)+\frac{(m-1)}{2N}W(\mu)+\frac{(m-1)}{N}\int
f\,d\mu
\]
and $L=\inf\{k(x,a)+f(x):a,x\in S_{\mu}^{*}\}$. The rest of the
arguments in the proof of $(\ref{eq29general})$ are analogous to
those used to justify $(\ref{eq29})$. \hfill $\Box$

\begin{lemma}\label{Lema1}
Let $p\geq 2$ and $p-2\leq s< p$. Assume that
$A\subset\mathbb{R}^{p}$ is closed and $f$ satisfies the
conditions $(\ref{conda})$ and $(\ref{condb})$ $($or
$(\ref{condb2})$ in the case $p=2$, $s=0$$)$. Let
$\lambda=\lambda_{s,f}$ be the equilibrium measure solving the
Gauss variational problem on $A$ in the presence of $f$. Then
\begin{itemize}
\item[(i)] for any measure
$\nu\in\mathcal{M}_{1}(\mathbb{R}^{p})$,
\begin{equation}\label{eq:ecuacion1}
``\inf_{x\in S_{\lambda}}"\,(U^{\nu}_{s}(x)+f(x))\leq
W_{s,f}(\lambda)\,,
\end{equation}
where $S_{\lambda}$ denotes the support of $\lambda$, and
$``\inf"$ means that the infimum is taken quasi-everywhere.

\item[(ii)] If $\nu\in\mathcal{M}_{1}(A)$, then
\begin{equation}\label{eq:ecuacion2}
``\sup_{x\in S_{\nu}}"\,(U^{\nu}_{s}(x)+f(x))\geq
W_{s,f}(\lambda)\,,
\end{equation}
where $S_{\nu}$ denotes the support of $\nu$, and $``\sup"$ means
that the supremum is taken quasi-everywhere.

\item[(iii)] Suppose that $\nu\in\mathcal{M}_{1}(A)$ has finite
$s$-energy and there exists a constant $M$ such that
$U^{\nu}_{s}(x)+f(x)=M$ for q.e. $x\in S_{\nu}$ and
$U^{\nu}_{s}(x)+f(x)\geq M$ for all $x\in A$. Then $\nu=\lambda$
and $M=W_{s,f}(\lambda)$.
\end{itemize}
\end{lemma}
\begin{proof}
The case $p=2$, $s=0$ of this result is part of Theorems I.3.1 and
I.3.3 in \cite{SaffTotik}. We first justify (\ref{eq:ecuacion1}).
To the contrary, suppose that there exists a measure
$\nu\in\mathcal{M}_{1}(\mathbb{R}^{p})$ and a constant
$C>W_{s,f}(\lambda)$ such that
\[
U^{\nu}_{s}(x)+f(x)\geq C\qquad \mbox{for}\,\,\mbox{q.e.}\quad
x\in S_{\lambda}\,.
\]
From $(\ref{Gaussvarineq1Riesz})$ we obtain that
\begin{equation}\label{desig1}
U^{\lambda}_{s}(x)+C-W_{s,f}(\lambda)\leq U_{s}^{\nu}(x)\qquad
\mbox{for}\,\,\mbox{q.e.}\quad x\in S_{\lambda}\,.
\end{equation}
Since $I_{s}(\lambda)$ is finite, $S_{\lambda}$ is a compact set
with positive $s$-capacity. Therefore, there exists a unique
measure $\mu_{\lambda}\in\mathcal{M}_{1}(S_{\lambda})$ such that
$I_{s}(\mu_{\lambda})=w_{s}(S_{\lambda})>0$. Since
$U^{\mu_{\lambda}}_{s}\leq w_{s}(S_{\lambda})$ on
$\supp(\mu_{\lambda})$, applying the first maximum principle (cf.
\cite[Theorem 1.10]{Landkof}) it follows that
$U^{\mu_{\lambda}}_{s}\leq w_{s}(S_{\lambda})$ everywhere in
$\mathbb{R}^{p}$. Using (\ref{Gaussvarineq2Riesz}) we conclude
that $U^{\mu_{\lambda}}_{s}=w_{s}(S_{\lambda})$ q.e. on
$S_{\lambda}$.

If we define now the measure
$\eta:=(C-W_{s,f}(\lambda))\,w_{s}(S_{\lambda})^{-1}\mu_{\lambda}$,
(\ref{desig1}) yields
\begin{equation}\label{desg2}
U^{\lambda+\eta}_{s}(x)\leq U_{s}^{\nu}(x)\qquad
\mbox{for}\,\,\mbox{q.e.}\quad x\in S_{\lambda}\,.
\end{equation}
Since $\lambda$ and $\eta$ have finite energy, this inequality
holds $(\lambda+\eta)$-almost everywhere. Applying Theorem 1.27
(case $s=p-2$) and Theorem 1.29 (case $p-2<s<p$) from
\cite{Landkof} we obtain that the inequality (\ref{desg2}) holds
everywhere in $\mathbb{R}^{p}$. Finally, multiplying both sides by
$|x|^{s}$ and letting $|x|\rightarrow\infty$ it follows that
$C-W_{s,f}(\lambda)\leq 0$, which contradicts our initial
assumption.

Now we prove (\ref{eq:ecuacion2}). Let $L:=``\sup"_{x\in
S_{\nu}}(U^{\nu}_{s}(x)+f(x))$ and assume that $L$ is finite. It
follows from this assumption that $\nu$ has finite $s$-energy.
Using $(\ref{Gaussvarineq2Riesz})$ we have
\begin{equation}\label{eq:ecuacion3}
U^{\nu}_{s}(x)+W_{s,f}(\lambda)-L\leq U^{\lambda}_{s}(x)\qquad
\mbox{for}\,\,\mbox{q.e.}\quad x\in S_{\nu}\,.
\end{equation}
The same argument employed above to prove part (i) shows that
$W_{s,f}(\lambda)-L\leq 0$.

Finally, the assumptions of (iii) imply that
\[
``\inf_{x\in A}"\,(U^{\nu}_{s}(x)+f(x))=M=``\sup_{x\in
S_{\nu}}"\,(U^{\nu}_{s}(x)+f(x))\,,
\]
and consequently we obtain using $(\ref{eq:ecuacion1})$ and
$(\ref{eq:ecuacion2})$ that $M=W_{s,f}(\lambda)$. Taking $C=M$ in
$(\ref{desig1})$ and $L=M$ in $(\ref{eq:ecuacion3})$ we conclude
that $U_{s}^{\nu}=U_{s}^{\lambda}$ everywhere in $\mathbb{R}^{p}$,
which implies that $\lambda=\nu$ by Theorem 1.15 from
\cite{Landkof}.
\end{proof}

\noindent{\textsc{Proof of Lemma \ref{lemanBW}.}} From
(\ref{eq:BWL1}) and Lemma \ref{Lema1}(i) applied to the measure
$\nu:=(1/n)\sum_{i=1}^{n}\delta_{x_{i}}$ we obtain that
$W_{s,f}(\lambda)\geq M/n$. Using (\ref{Gaussvarineq1Riesz}) and
(\ref{eq:BWL1}) we have
\begin{equation}\label{eq:pruebaBWL1}
U^{\nu}_{s}(x)+W_{s,f}(\lambda) -\frac{M}{n}\geq
U^{\lambda}_{s}(x)\qquad \mbox{for}\,\,\mbox{q.e.}\quad x\in
\supp(\lambda)\,.
\end{equation}
The same argument employed to prove Lemma \ref{Lema1}(i) shows
that the inequality $(\ref{eq:pruebaBWL1})$ is valid everywhere in
$\mathbb{R}^{p}$, which is precisely $(\ref{eq:BWL2})$. Finally,
$(\ref{eq:BWL3})$ is a consequence of $(\ref{eq:BWL2})$ and
$(\ref{Gaussvarineq2Riesz})$. \hfill $\Box$

\vspace{0.2cm}

\noindent{\textsc{Proof of Corollary \ref{corolario}.}} The fact
that $a_{n}$ is well-defined for all $n\geq 1$ follows from
conditions $(\ref{conda})$ and $(\ref{condb})$ (or
$(\ref{condb2})$ in the case $p=2$, $s=0$). Applying Lemma
$\ref{lemanBW}$ to
$\{x_{1},\ldots,x_{n}\}:=\{a_{1},\ldots,a_{n}\}$ and
\[
M:=\sum_{i=1}^{n}\frac{1}{|a_{n+1}-a_{i}|^{s}}+n f(a_{n+1})\,,
\]
it follows that $a_{n}\in S_{\lambda}^{*}$ for all $n\geq 2$. The
case $p=2$, $s=0$ is justified in the same way. It is clear from
the proof of Theorem $\ref{theoexternalgreedy}$ that
$(\ref{asympweight})$-$(\ref{eq29})$ are valid for the weighted
greedy $f$-energy sequence $(a_{n})_{n=1}^{\infty}$. \hfill $\Box$

\vspace{0.2cm}

\noindent{\textsc{Proof of Corollary \ref{corolario2}.}} For every
$N\geq 0$, the existence of the minimizing configuration
$\{a_{mN+1},\ldots,a_{m(N+1)}\}$ is guaranteed by the conditions
$(\ref{conda})$ and $(\ref{condb})$ (or $(\ref{condb2})$ in the
case $p=2$, $s=0$).

Next, we show that
$\omega_{N}:=\{a_{mN+1},\ldots,a_{m(N+1)}\}\subset
S_{\lambda}^{*}$ for every $N\geq 0$. It follows from the
definition of $\omega_{N}$ that for each $i\in\{1,\ldots,m\}$, the
inequality
\begin{equation}\label{eq:prueba2}
\sum_{l=1}^{mN}\frac{1}{|a_{mN+i}-a_{l}|^{s}}+\sum_{j=1,j\neq
i}^{m}\frac{1}{|a_{mN+i}-a_{mN+j}|^{s}}+((N+1)m-1)f(a_{mN+i})
\end{equation}
\[
\leq \sum_{l=1}^{mN}\frac{1}{|x-a_{l}|^{s}}+\sum_{j=1,j\neq
i}^{m}\frac{1}{|x-a_{mN+j}|^{s}}+((N+1)m-1)f(x)
\]
holds for all $x\in A$. (If $N=0$ then the first term on both
sides of the inequality doesn't appear in the expression.) If we
denote the left hand side of $(\ref{eq:prueba2})$ by $M$, and
apply Lemma $\ref{lemanBW}$ to
$\{x_{1},\ldots,x_{(N+1)m-1}\}=\{a_{l}\}_{l=1}^{mN}\cup\{a_{mN+j}\}_{j=1,j\neq
i}^{m}$, then (\ref{eq:BWL2}) implies that $a_{mN+i}\in
S_{\lambda}^{*}$.

It is clear that the sequence $(a_{n})_{n\geq 1}$ is a weighted
greedy $(m,f,\lambda)$-energy sequence and, therefore, all the
assertions of Theorem \ref{theogreedygeneral} are applicable to
$(a_{n})_{n\geq 1}$.\hfill $\Box$

\vspace{0.2cm}

\noindent{\textsc{Proof of Proposition \ref{propcircweight}.}} It
is easy to see that
\[
\int_{S^{p-1}}\frac{1}{|r\overline{y}-x|^{p-2}}d\sigma_{p-1}(\overline{y})=\left\{
\begin{array}{ccc}
1/r^{p-2}, & \mbox{if} & |x|\leq r\,, \\
\\
1/|x|^{p-2}, & \mbox{if} & |x|>r\,.
\end{array}
\right.
\]
Let $\nu$ be the measure supported on
$\{x\in\mathbb{R}^{p}:r_{0}\leq |x|\leq R_{0}\}$ whose expression
is given by the right-hand side of $(\ref{eq:eqcircweight})$. From
the definition of $r_{0}$ and $R_{0}$ it follows that $\nu$ is a
probability measure and by simple computations we obtain that the
potential $U^{\nu}_{p-2}$ coincides with the function on the
right-hand side of $(\ref{eq:extfieldcirc})$. Therefore
\begin{equation}\label{eq:igualdad}
U^{\nu}_{p-2}(x)+f(x)=\frac{1}{R_{0}^{p-2}}+f(R_{0})\,,\qquad
r_{0}\leq |x|\leq R_{0}\,.
\end{equation}
Applying the definitions of $r_{0}$ and $R_{0}$ again, we get that
$f(|x|)\geq f(r_{0})$ if $|x|\leq r_{0}$ and
$f(|x|)+1/|x|^{p-2}\geq f(R_{0})+1/R_{0}^{p-2}$ if $|x|\geq R_{0}$
(regarding $f$ as a function of $\mathbb{R}_{+}$). As a
consequence
\begin{equation}\label{eq:desigualdad}
U^{\nu}_{p-2}(x)+f(x)\geq \frac{1}{R_{0}^{p-2}}+f(R_{0})
\end{equation}
for all $x\in\mathbb{R}^{p}$. Therefore, it follows from
$(\ref{eq:igualdad})$, $(\ref{eq:desigualdad})$, and Lemma
\ref{Lema1}, that $\nu=\lambda_{p-2,f}$ and
$(\ref{eq:wflambdacirc})$ holds.\hfill $\Box$

\vspace{0.2cm}

\noindent{\bf Acknowledgments.} I take this opportunity to thank
my research advisor, Professor Edward Saff, for his dedication and
support, and for the many helpful discussions we held on these and
other related topics. I am also grateful to Natalia Zorii for her
careful reading of this manuscript and her suggestions to improve
it.

\bibliographystyle{amsalpha}

\providecommand{\bysame}{\leavevmode\hbox to3em{\hrulefill}\thinspace}
\providecommand{\MR}{\relax\ifhmode\unskip\space\fi MR }
\providecommand{\MRhref}[2]{%
  \href{http://www.ams.org/mathscinet-getitem?mr=#1}{#2}
}
\providecommand{\href}[2]{#2}
\begin{thebibliography}{}

\end{thebibliography}


\begin{thebibliography}{A}

\bibitem{Choquet2}
G. Choquet, \textit{Theory of capacities}, Ann. Inst. Fourier
\textbf{5} (1955), 131--295.

\bibitem{Choquet}
G. Choquet, \textit{Diam\`{e}tre transfini et comparaison de
diverses capacit\'{e}s}, Technical report, Facult\'{e} des
Sciences de Paris (1958).

\bibitem{Edrei}
A. Edrei, \textit{Sur les d\'{e}terminants r\'{e}currents et les
singularit\'{e}s d'une fonction donn\'{e}e par son
d\'{e}veloppement de Taylor}, Compositio Mathematica \textbf{7}
(1939), 20--88.

\bibitem{FarkasNagy}
B. Farkas and B. Nagy, \textit{Transfinite diameter, Chebyshev
constant and energy on locally compact spaces}, Potential Anal.
\textbf{28} (2008), 241--260.

\bibitem{Fuglede}
B. Fuglede, \textit{On the theory of potentials in locally compact
spaces}, Acta Math. \textbf{103} (1960), 139--215.

\bibitem{Landkof}
N.S. Landkof, \textit{Foundations of Modern Potential Theory},
Springer-Verlag, Heidelberg, 1972.

\bibitem{Leja}
F. Leja, \textit{Sur certaines suites li{\'e}e aux ensembles plans
et leur application {\`a} la repr{\'e}sentation conforme}, Ann.
Polon. Math. \textbf{4} (1957), 8--13.

\bibitem{LS}
A. L\'{o}pez Garc\'{i}a and E.B. Saff, \textit{Asymptotics of
greedy energy points}, to appear in Math. Comp.

\bibitem{MhaskarSaff}
H.N. Mhaskar and E.B. Saff, \textit{Where does the sup norm of a
weighted polynomial live? $($A generalization of incomplete
polynomials$)$}, Constr. Approx. \textbf{1} (1985), 71--91.

\bibitem{MhaskarSaff2}
H.N. Mhaskar and E.B. Saff, \textit{Weighted analogues of
Capacity, Transfinite Diameter, and Chebyshev Constant}, Constr.
Approx. \textbf{8} (1992), 105--124.

\bibitem{Ohtsuka}
M. Ohtsuka, \textit{On potentials in locally compact spaces}, J.
Sci. Hiroshima Univ., ser. \textbf{A 1} (1961), 135--352.

\bibitem{SaffTotik}
E.B. Saff and V. Totik, \textit{Logarithmic Potentials with
External Fields}, Grundlehren der Mathematischen Wissenschaften,
Vol. 316, Springer-Verlag, 1997.

\bibitem{Siciak}
J. Siciak, \textit{Two criteria for the continuity of the
equilibrium Riesz potentials}, Ann. Soc. Math. Polonae (1970),
91--99.

\bibitem{Zorii}
N.V. Zorii, \textit{Equilibrium problems for potentials with
external fields}, Ukrainian Math. J. \textbf{55} (2003),
1588--1618.

\bibitem{Zorii2}
N.V. Zorii, \textit{Equilibrium potentials with external fields},
Ukrainian Math. J. \textbf{55} (2003), 1423--1444.

\end{thebibliography}

\end{document}